\documentclass[sigplan,screen]{acmart}

\usepackage{listings}
\usepackage{enumerate}
\usepackage{enumitem}
\usepackage{xspace}
\usepackage{url}
\usepackage{pgfplots}\pgfplotsset{compat=1.18}
\usepackage{hyperref}
\usepackage{cleveref}
\usepackage{xspace}
\usepackage{microtype}
\microtypesetup{protrusion=false}

\usepackage{pmboxdraw}      
\usepackage{newunicodechar} 
\newunicodechar{¬}{\ensuremath{\neg}}
\newunicodechar{∧}{\ensuremath{\wedge}}
\newunicodechar{∨}{\ensuremath{\vee}}

\usepackage{xcolor}

\newcommand{\stitle}[1]{\noindent\textbf{#1\xspace}}
\newcommand{\systemname}[0]{\texttt{FlowCheck}\xspace}

\newsavebox{\promobox}

\usepackage{stmaryrd} 
\newcommand{\tok}[1]{\text{\ttfamily #1}}   
\newcommand{\pq}[1]{\text{\sffamily #1}}    
\newcommand{\handler}{\mathit{handler}}
\newcommand{\Tsem}[1]{\mathsf{T}\llbracket #1 \rrbracket}
\newcommand{\Csem}[1]{\mathsf{C}\llbracket #1 \rrbracket}
\newcommand{\Vsem}[1]{\mathsf{V}\llbracket #1 \rrbracket}
\newcommand{\Esem}[2]{\mathsf{E}\llbracket #2 \rrbracket_{#1}}
\newcommand{\Rsem}[2]{\mathsf{R}\llbracket #2 \rrbracket_{#1}}

\lstdefinelanguage{JavaScript}{
  keywords={const, let, var, typeof, new, true, false, catch, function, return, null, switch, if, in, while, do, else, case, break, parseFloat},
  keywordstyle=\bfseries\color{blue},
  identifierstyle=\color{black},
  sensitive=false,
  comment=[l]{//},
  morecomment=[s]{/*}{*/},
  commentstyle=\itshape\color{gray},
  stringstyle=\color{red},
  morestring=[b]',
  morestring=[b]"
}

\setcopyright{cc}
\setcctype{by}
\acmDOI{10.1145/3843750.3843844}
\acmYear{2026}
\copyrightyear{2026}
\acmISBN{979-8-4007-2986-7/2026/10}
\acmConference[LMPL '26]{Proceedings of the 2nd ACM SIGPLAN International Workshop on Language Models and Programming Languages}{October 4--9, 2026}{Oakland, CA, USA}
\acmBooktitle{Proceedings of the 2nd ACM SIGPLAN International Workshop on Language Models and Programming Languages (LMPL '26), October 4--9, 2026, Oakland, CA, USA}
\acmSubmissionID{splashws26lmplmain-p81-p}
\received{2026-07-11}
\received[accepted]{2026-08-14}

\begin{document}

\title{FlowCheck: Helping End-Users Specify and Verify Intent in Vibe-Coded Web Apps}

\author{Reya Vir}
\orcid{0009-0003-3117-5655}
\affiliation{%
  \institution{\raisebox{-0.20em}{\includegraphics[height=1.2em]{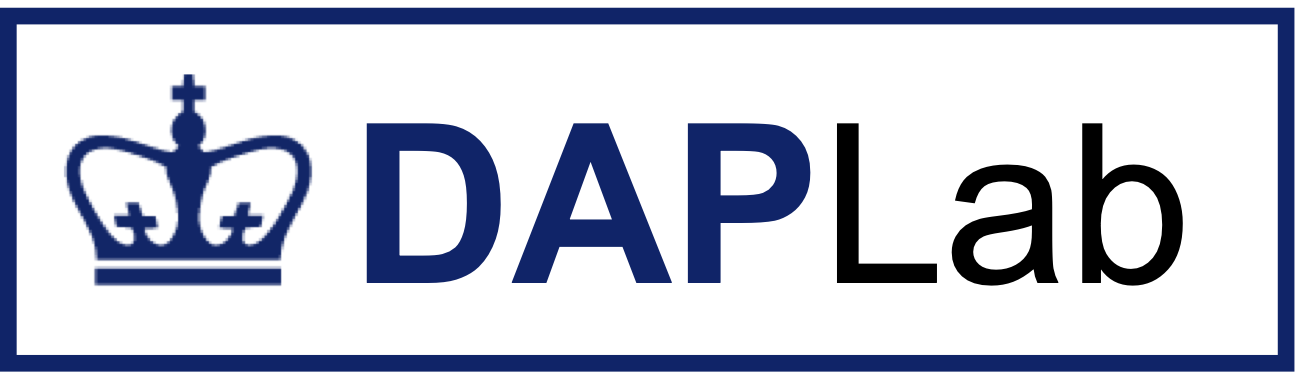}} Columbia University}
  \city{New York}
  \country{USA}
}
\email{reyavir@cs.columbia.edu}

\author{Lydia Chilton}
\orcid{0000-0002-1737-1276}
\affiliation{%
  \institution{\raisebox{-0.20em}{\includegraphics[height=1.2em]{images/daplab_logo.png}} Columbia University}
  \city{New York}
  \country{USA}
}
\email{chilton@cs.columbia.edu}

\author{Zhuo Zhang}
\orcid{0000-0002-6515-0021}
\affiliation{%
  \institution{\raisebox{-0.20em}{\includegraphics[height=1.2em]{images/daplab_logo.png}} Columbia University}
  \city{New York}
  \country{USA}
}
\email{zz3474@columbia.edu}

\author{Eugene Wu}
\orcid{0000-0003-4254-6688}
\affiliation{%
  \institution{\raisebox{-0.20em}{\includegraphics[height=1.2em]{images/daplab_logo.png}} Columbia University}
  \city{New York}
  \country{USA}
}
\email{ewu@cs.columbia.edu}

\begin{abstract}
Vibe-coded applications often contain silent behavioral failures in which the interface appears functional even though user-visible information does not flow to the expected state or output. We introduce \systemname, a constraint language to specify these user-visible information flows directly through the application interface, where constraints can also be displayed and inspected without reading code, and are structured enough for reliable LLM generation. \systemname translates the constraints into deterministic CodeQL analyses, and we evaluate it across four applications generated via Claude Code, and compare with three coding models as bug-finding baselines. We find that \systemname correctly translates and flags all 30 of our injected constraint violations with no false positives. In contrast, frontier models (Claude Opus 4.7, DeepSeek V3, and Gemini Pro) showed significantly lower accuracy when prompted to find bugs in the same code, with none achieving full accuracy. This approach lets vibe coders state intent in terms of the interface they understand, and checks it deterministically against the code they do not.
\end{abstract}



\maketitle

\section{Introduction}

Large language model (LLM) coding agents have democratized software development. End users with little programming experience can now ``Vibe code'' complex applications and iterate on their features via natural language. These {\it Vibe Coders} evaluate progress primarily by interacting with the application rather than inspecting its implementation. This shifts the bottleneck from writing code to determining whether the generated application behaves as intended.

Coding agents remain unreliable. They may misunderstand requests, omit necessary state updates, introduce incorrect data flows, or break previously working behavior. Prior work has documented incomplete implementations, regressions, and incorrect agent-generated code \cite{dente2026constraintdecayfragilityllm, vero2025baxbenchllmsgeneratecorrect}. Beyond code-level errors, large-scale analysis of real coding-agent sessions shows that agents frequently break down on what users actually want~\cite{tang2026codingagentsfailusers}. These \textbf{\textit{Silent Behavioral Failures}}---the application runs but its observable behavior violates user intents---are especially difficult to detect when the application compiles, renders normally, and produces plausible feedback. A button may report success even though no data was saved, or a displayed value may update without reflecting the state it is supposed to represent. 

Existing debugging techniques like unit, integration, and end-to-end testing can validate behavior, but require users to identify test cases and encode the expected result as an executable oracle. LLM-based testing and debugging reduce the authoring burden, but we find these are probabilistic and limited in the types of failures they can identify. Traditional static analysis and program verification techniques are deterministic for predefined program properties, but are designed for programmers rather than vibe coders.

Vibe coders therefore need a way to express what they expect in terms of the interface they understand and to deterministically check whether the implementation supports that behavior. We focus on web applications, where user intents can be expressed as constraints over flows between user actions, interface-level updates, persistent state, and backend calls. For example, a user may expect that clicking a button should update a displayed total based on an input, or that submitting a form will persist its contents.

We introduce \systemname, a system that expresses and checks data and control flow constraints over visible interface elements and application effects---for instance, that an action should not update a particular component, values should derive from specific inputs, state must persist across reloads, or an action must trigger an API call.  
We define a constraint language that users can express and visualize directly in the application interface, and that can be easily generated by LLMs. The constraints compile to static analysis queries (CodeQL) over the application's event handlers, control and data flow, and interface and storage accesses.

\begin{example}\it
The user asks an agent to add a promotional-code feature to a vibe-coded shopping application. The agent creates a promo input and an ``Apply'' button. When the user enters a valid code, the interface displays ``Discount applied!'' even though the new total is neither persisted nor shown:

\begin{lrbox}{\promobox}
\begin{lstlisting}[
language=JavaScript,
basicstyle=\scriptsize\ttfamily,
keywordstyle=\bfseries,
commentstyle=\itshape,
frame=none
]
function applyPromo() {
 const getById = id => document.getElementById(id);
 const code = getById('promo-input').value;

 if (code === 'SAVE20') {
  // read and update cart state
  let tot = parseFloat(localStorage.getItem('cartTotal'));
  tot *= 0.8;

  // Update success message in UI
  getById('promo-msg').innerText = 'Discount applied!';
 
  // BUG: did not persist nor display updated total
  // localStorage.setItem('cartTotal', tot); // missing
 }
}
\end{lstlisting}
\end{lrbox}
\begin{center}\usebox{\promobox}\end{center}
\end{example}

Although the app runs and the message suggests the feature works, it violates the user's expectation that applying the code changes the cart total, even on refresh. In \systemname, users can specify that clicking the button should cause the promo code to update the displayed and persisted total. The analysis identifies that the message is updated but the discounted value does not flow to either destination.

Unlike LLM-as-judge approaches, \systemname returns the same result for  identical code and constraints, without executing the app. Here, our focus is on properties that can be statically analyzed, as they are lightweight and can be used within agent loops. We find that constraints are useful targets for agents to generate to aid testing their code, and violations provide details of the failure and intended behavior, which guides debugging and avoids relying on vibe-coder guesses. Our paper makes the following contributions:

\begin{enumerate}[nosep, leftmargin=1em]
\item We develop a taxonomy of silent behavioral failures from a formative study spanning four applications, three coding models, and four iteration steps. From these, we identified 23 distinct failures, where nearly half (48\%) can be expressed and checked fully using our static constraints.
\item We introduce a constraint language and interface through which end users express expected behavior using visible elements, actions, and effects.
\item We present a deterministic static-analysis system that translates these constraints into CodeQL queries and checks the required control-flow and data-flow relationships.
\item We evaluate \systemname on 30 injected failures across four web applications (14--21 constraints each), where it achieves 100\% detection accuracy, compared to at most 26/30 (87\%) for the best LLM baseline (Claude, best prompt). We further show that agents can author valid constraints in our language, and that using them to check code raises the agents' bug-catch rate on a test app, an early sign the language is learnable and usable by LLMs.

\end{enumerate}

\section{Related Work}

\subsection{Evaluating LLM-Generated Applications}

Coding agents struggle as requirements iteratively grow, across front-end development~\cite{10.1145/3786335.3813180} and backend correctness constraints~\cite{dente2026constraintdecayfragilityllm,vero2025baxbenchllmsgeneratecorrect}. We focus on \emph{silent behavioral failures}: the application runs and presents plausible output, but user-visible information does not flow to the expected state or  element. This differs from prior semantic failures~\cite{246326,280920} by grounding correctness in end user's expectations.

LiveCodeBench and DebugBench evaluate self-contained problems with predefined input-output oracles~\cite{jain2024livecodebenchholisticcontaminationfree,tian2024debugbenchevaluatingdebuggingcapability}, while SWE-bench evaluates patches to existing repositories~\cite{jimenez2024swebenchlanguagemodelsresolve}. These benchmarks therefore do not capture expectations expressed through a newly generated application's interface. Recent vibe-coding benchmarks evaluate complete applications using browser agents or LLM judgments~\cite{10.1145/3786335.3813180,bansal2026vibepassvibecodersreally}. However, recent position work argues that vibe coding needs more deterministic checking~\cite{10.1145/3759425.3763390}; our formative study characterizes failures, while \systemname lets users state and deterministically check the expected information flows.

Automated test generation is another approach to check agent-generated code, but faces the oracle problem, where a test is dependent on knowing what the correct behavior should be~\cite{6963470}. Tools such as EvoSuite~\cite{10.1145/2025113.2025179} are able to produce test sets and suggest assertions, but those assertions are based on what the current code does, rather than what it should do. The same issue occurs with LLM-based test generators like TestPilot~\cite{schäfer2023empiricalevaluationusinglarge} which generates JavaScript (JS) unit tests by prompting a model with the implementation of the function, so its oracles may reflect the existing code. If the code contains silent failures, the generated oracle may not be able to detect it. In our approach, our constraints supply the oracle from an outside user or agent, representing the user's intended behavior separate from implementation.

\subsection{Specifications and LLM Debugging}
SpecGen, AutoSpec, Clover, and PATAgent use LLMs to generate or formalize specifications and then apply verification tools~\cite{ma2025specgenautomatedgenerationformal,wen2024enchantingprogramspecificationsynthesis,sun2024cloverclosedloopverifiablecode,zuo2025patagentautoformalizationmodelchecking}. Their specifications describe code using formal abstractions that non-programmers cannot easily inspect. Our language is intentionally less expressive: it describes user-visible actions, elements, and information flows, can be authored and displayed within the application interface, and is simple enough for an LLM to generate.

LLM-based debugging and repair instead ask a model to diagnose or correct the program. This has been done in various ways: by prompting it to explain and revise its own code~\cite{chen2023teachinglargelanguagemodels}, by using execution traces or semantic context~\cite{wang2025codesemanticshelpcomprehensive}, or more recently by treating the LLM as an autonomous agent that plans and invokes tools~\cite{10.1109/ICSE55347.2025.00157}. However, self-repair gives inconsistent gains, and models are limited by their ability to provide actionable feedback on why code fails ~\cite{olausson2024selfrepairsilverbulletcode}. Even where repair succeeds, it does not leave the user with a persistent, independently checkable statement of the intended behavior. Our constraints remain visible to the user, and their analysis is deterministic regardless of whether the user or an LLM authored them.

\subsection{End-User Web Testing and Programming}

Dynamic web-testing systems such as GUITAR, Crawljax, and Atusa execute interaction sequences and check resulting interface states or DOM invariants~\cite{10.1007/s10515-013-0128-9,10.1145/2109205.2109208,5728834}. Quickstrom similarly checks user-facing behavior, but it uses a runtime approach, with specifications written in a temporal logic aimed at web programmers~\cite{10.1145/3519939.3523728}. Our constraints similarly describe application-level behavior, but are checked statically as information-flow relationships without executing or crawling the application. This makes checks immediate and repeatable, but excludes runtime-only properties such as external API results or dynamically generated code.

End-user programming systems already make the program and state visible to the user.  Users can debug by directly asserting e.g., spreadsheet values, select interface artifacts, or ask questions about program behavior~\cite{1201191,302163.302183,10.1145/985692.985712}. Trigger-action programming (TAP) further shows that \texttt{if--then} rules can be accessible to non-programmers~\cite{10.1145/3411764.3445567}. 

Recent work shows that natural language is difficult for users and programmers to state their intent~\cite{10.1145/3706598.3713271}. Vibe coders  struggle with reading and evaluating correctness of LLM-generated code~\cite{10.1145/3613904.3642706}. 
Prior work frames this as an abstraction gap between user intent and generated code, and bridges this by making the generated code legible or visualized so the user can form an accurate mental model~\cite{10.1145/3544548.3580817}. We instead start with the visible interface, and define data- and control-flow constraints from  what users can directly express. Formally stating the intended behavior lets \systemname mechanically check and enforce them on behalf of the vibe coder.  

\section{Formative Study: Vibe Coding Failures}
We conducted a formative study to characterize failures that arise during iterative vibe coding. Across four applications and three coding systems, we observed 23 distinct failures. Every application and system exhibited at least one failure.

\subsection{Methodology}
We used a closed-weight model (Claude Opus 4.7), an open-weight model (DeepSeek V3), and a commercial vibe-coding system (Lovable). For each system, we generated four applications with four scripted stages: initial generation, feature addition, refactoring, and feature modification. We fixed the prompts across systems, requested plain JavaScript without frameworks, saved the code after each stage, and manually checked it against expectations written before generation.

The applications were a workshop speaker scheduler; a column-based task board refactored into a whiteboard with dependencies between tasks; a shopping application with promo codes and limited-edition items; and an image gallery with favorites, albums, navigation, and a public image API.

\subsection{Observed Failures}
\label{sec:formative-study-failures}

We grouped the 23 failures into five recurring categories.

\smallskip
\noindent
\textbf{Surface-level correctness}. The visible output is disconnected from the data or action it claims to represent. In the image galleries produced by Claude, DeepSeek, and Lovable, clicking a thumbnail opened a different image; data should flow from the clicked thumbnail to the popup. In DeepSeek's shopping app, the \texttt{VIP20} promo code failed when the cart contained limited-edition items; the displayed total ignored the promo input and cart contents.

\smallskip
\noindent
\textbf{Dropped constraints across iterations.}
A relationship established in one iteration, disappeared after a later modification. All three systems dropped the rule that a speaker could occupy only one slot after a subsequent prompt allowed multiple speakers per slot. Similarly, after task boards became whiteboards, Claude, DeepSeek, and Lovable allowed a task to move directly from ``not started'' to ``finished,'' despite an earlier rule forbidding that state update. In both cases, the interface remained functional, but an earlier restriction on which actions could update state was lost.

\smallskip
\noindent
\textbf{Missing implied functionality.}
The interface advertises an action without connecting it to its corresponding effect. The Claude and DeepSeek task boards displayed an archive area labeled ``Move a card here to set aside,'' but dropping a card did not update the archive; only a separate button worked. DeepSeek's shopping app presented Tinder-style cards, but swiping did not advance the displayed item and users had to fall back to buttons. These failures expose missing action-to-effect links, although gesture semantics such as dragging and swiping may need runtime support beyond static analysis.

\smallskip
\noindent
\textbf{Breaking changes after modification.}
A new feature or refactoring severed a relationship that previously worked. After the task board was converted to a whiteboard, cards from all three systems no longer displayed their status: the underlying task state no longer flowed to the visible card. In DeepSeek's dependency editor, adding $A\rightarrow B$ and then $A\rightarrow C$ silently rewrote the graph as $A\rightarrow C\rightarrow B$, so the second action overwrote rather than preserved the relationship.

\smallskip
\noindent
\textbf{Incomplete persistence.}
State was not saved, restored, or associated with the correct data. In the image gallery, favorites were stored by position rather than image identity, so reloading caused saved favorites to refer to different images (Claude, DeepSeek, Lovable). In Lovable's speaker scheduler, neither the speaker list nor the schedule survived a reload, despite the prompt describing the generated people as ``my speaker list.'' The first failure persisted the wrong source; the second omitted the save-and-restore path entirely.

\smallskip
\noindent
\textbf{Failures in the Wild.} Beyond our study, we observed the same patterns in public vibe-coding transcripts\footnote{\url{https://github.com/jennygzma/vibecoding-chatqa}}: one agent generated search and history interfaces without creating the required database tables, while another left a game permanently displaying ``loading'' because an asynchronous result never reached the interface. 

\subsubsection{Summary of Findings}
We find that 11 of the 23 failures share a common structure: missing or incorrect relationships between a user action and an observable effect. These included whether a component is updated, whether its value derives from the correct inputs, whether an update is prohibited under a condition, and whether state persists across reloads. Because these relationships span event handlers, control flow, DOM updates, and storage operations, they are non-local and difficult to reason about from code fragments, yet visible to the user. This motivates an interface-grounded constraint language that makes such information flows explicit and mechanically checkable.
\section{System Overview}
\label{sec:overview}
\begin{figure}[h]
    \centering
    \includegraphics[width=0.475\textwidth]{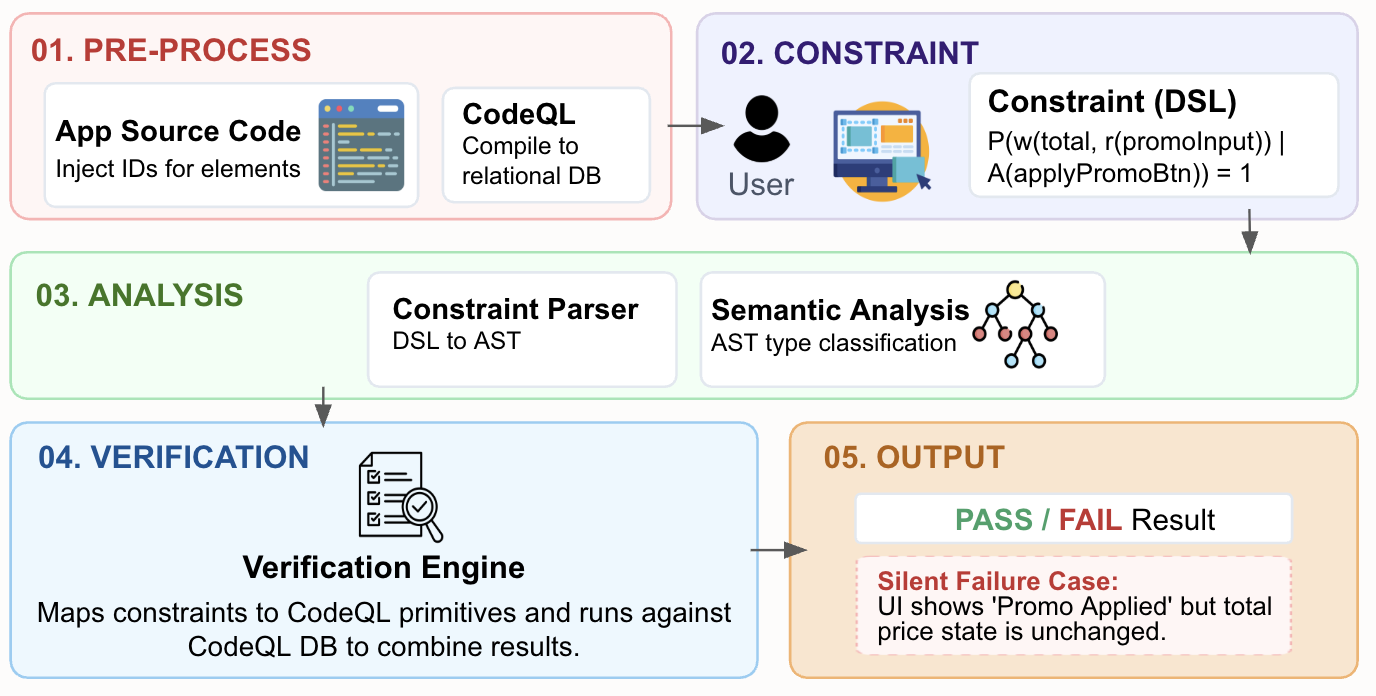}
    \caption{\systemname Core Engine. The app source code is preprocessed by CodeQL to support later static analysis queries. User- or LLM- authored constraints are compiled into CodeQL primitives, and run to find violations. Finally, we output the constraints that passed or failed. \protect\footnotemark
    }
    \label{fig:system-overview}
\end{figure}
\footnotetext{Icons from \url{https://www.flaticon.com}.}

Our system helps users verify and debug their vibe coded web applications. Instead of asking agents to debug, users author constraints by clicking through the behavior they expect, and \systemname verifies if the code aligns with that behavior. As shown in Figure~\ref{fig:user-constraint-authoring}, users select interface elements, actions, and storage state as nouns and specify the flows between them. Thus, the user interacts with the app as normal, without needing to learn or inspect code.

The workflow has the following three steps: First, code is preprocessed to add element identifiers (Section~\ref{sec:preprocessing}). Second, users author constraints through an interface running their app, or an agent generates them (Sections~\ref{sec:ui} and~\ref{sec:authoring}). Third, constraints compile into static analysis queries that run against the code, and returns pass/fail results. (Section~\ref{sec:compiling-semantics}).

CodeQL is a static code analysis engine\footnote{\url{https://codeql.github.com/}} that compiles source code into a relational database that models the abstract syntax tree, control-flow graph, and data-flow graph. We pre-process HTML separately in order to identify interface elements (Section~\ref{sec:preprocessing}).
Thus, constraints reduce to queries over the database and pre-processed HTML;  each query result is an instance of a match (e.g., a code path) that violates some expected behavior.

\section{Constraint Language}
\label{sec:language}

\subsection{Design Principles}
Drawing from related work in end user programming, we made three design decisions:
\begin{itemize}[leftmargin=*]

    \item \textbf{Observability:} The end-user's mental model is the interface, and they reason about expectations in terms of visible UI states. For this reason, we restrict constraints to observable elements, or key components we make visible to the user (API, storage).
    \item \textbf{Action to behavior framing:} every constraint is conditioned on a single user action, motivated by the trigger-action structure. Users reason about intent in terms of cause and effect (TAP), where an action leads to (or does not lead to) an expected effect~\cite{10.1145/2556288.2557420}.
    \item \textbf{Static verifiability:} We allow users to express only things we can check statically. Constraints that would require running the app (such as confirming an actual value from an API call, or detecting a race condition) are out of scope, as they require runtime tracing. Static analysis ensures it is deterministic and does not vary across traces.
\end{itemize}

\subsection{Abstractions}
We make the following abstractions to express constraints over the app's behavior. 
These are designed to be easy for a user to understand, and also easy for an LLM to generate.

\smallskip \noindent \textbf{Components:} The components of these constraints are strictly things that can be read from or written to, such as UI elements, database tables, and APIs.

\smallskip \noindent \textbf{Actions:} User actions (e.g. clicking) and system actions (e.g API call, page load).

\smallskip \noindent \textbf{Reads and Writes:} 
$r()$, $w()$ over components. 
A write $w(c)$ means that there is a value written to component c; read $r(c)$ means that a value is read from component c.

\smallskip \noindent \textbf{Sources:} The sources of data for a component. This allows us to express a set of sources that a write must read from. For example,
$w(c, r(a)+1)$ requires c to be written with value $r(a)+1$, and $w(c, sources=\{r(a), r(b)\})$ requires c to be written with a value derived from exactly this set of reads.

Each constraint is a logical statement about whether an event occurs or not given a condition. The event either must always occur ($P=1$) as in all paths from the action must result in the expected event, while $P=0$ means this should never happen given the action. These map to our all paths and no paths checks over the code.

\subsection{Grammar}
Our grammar focuses on parsing probabilistic constraints of the form: 
$$P(event \mid action) = probability.$$
The {\it action} is the user or system event that the constraint is conditioned on, while the {\it event} is our expected outcome (e.g. write to a component or API call). It supports nested logic expressions (and/or/not/xor), user and system actions, reads/writes to components, and expected probabilities.

\begin{example}\it
For example, 
$$P(w(cartCount) \mid A(addCharger)) = 1,$$ 
means:
Every time the user takes action of clicking the add charger button, we expect that the cartCount component is written to every single time ($P=1$). This means that all execution paths from the addCharger event handler must write to the cartCount component with a probability of 1. 
\end{example}

\begin{figure}[h]
    \centering
\includegraphics[width=\columnwidth]{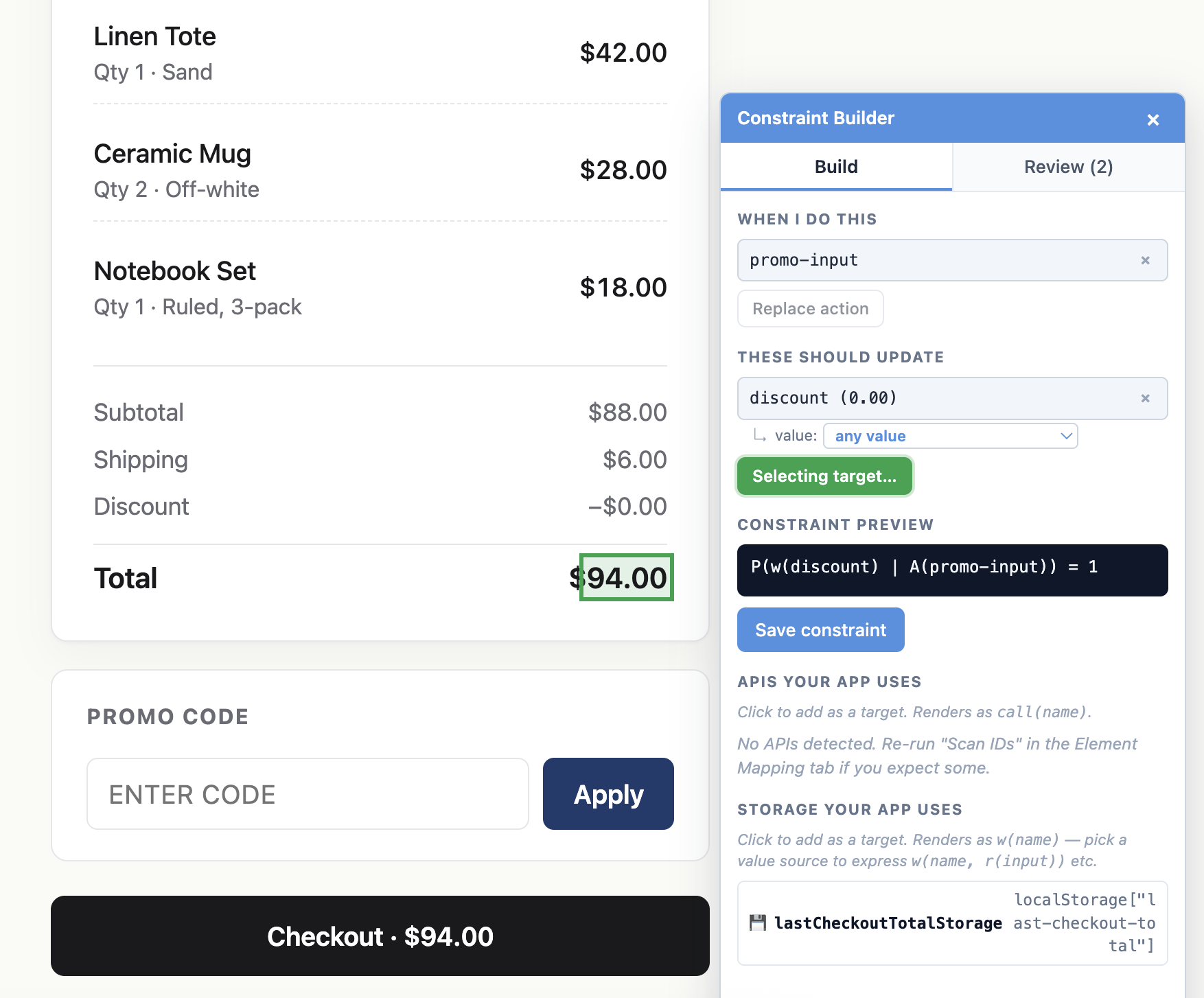}
    \caption{User authoring a constraint via the overlay. The user selects the promo input as the action and \texttt{discount} as a write target, then adds a second target (cart total, highlighted in green), to form a compound constraint. Users select visible components, without having to learn code or syntax.}
    \label{fig:user-constraint-authoring}
    \Description{Screenshot of a shopping cart page with the  overlay open beside it. The promo code input is highlighted as the selected action, and the discount and cart total fields are highlighted as write targets. The overlay panel shows the resulting compound constraint.}
\end{figure}

While most of the grammar are standard logical and arithmetic expressions, the types of state and atomic operations are worth mention. There are three types of state that serve as primitives that expressions are built on: UI elements that the user can see, storage entries for local or session storage, and APIs for external services including external storage. The grammar uses the \texttt{id} to reference every identifier position. During semantic analysis, each \texttt{id} is classified as a UI element, storage entry, or API from the mapping, which we use to ensure the constraint is valid. Each corresponds to one part of the authoring interface in Figure~\ref{fig:user-constraint-authoring}.

There are four atomic operations over this state: A write w() means a value is written to a component. It has optional constraints on the value or its sources. A user action A() names an event the user takes (e.g button clicks). A system event call() represents an API invocation. A persist event persist() represents state that survives across reloads.
The grammar permits actions and guards in any atom position, but our semantic analysis (Section~\ref{sec:analysis}) validates constraints and rejects those that put a user action or bare guard on the event side, or multiple actions on the condition side.

\begin{figure}[t]
\begin{lstlisting}[
  basicstyle=\ttfamily\small,
  frame=tb,
  columns=fullflexible,
  xrightmargin=2pt
]
constraint = "P(" lex "|" lex ")" "=" NUM EOF

lex  = lex (OR | AND | XOR) lex | NOT lex | atom
atom = write | action | call | persist | guard

write   = "w(" id ["," expr] ")" |
          "w(" id "," "sources=" srcset ")"
action  = "A(" id ")"                  // user action
call    = "call(" id ")"
persist = "persist(" id ")"
guard   = expr cmp expr | expr IN range

srcset  = "{" [srcitem {"," srcitem}] "}"
srcitem = "r(" id ")"

expr = expr ("+" | "-" | "*" | "/") expr |
       "r(" id ")" | lit

\end{lstlisting}
\caption{Constraint syntax in extended Backus--Naur form, a simplified subset of our grammar; see Appendix~\ref{sec:appendix-grammar} for the full grammar.}
\label{fig:constraint-syntax}
\end{figure}

\subsection{Expressiveness}
We enumerated the kinds of user-visible constraints a user might want to express about a web app, by drawing from our formative study (Section~\ref{sec:formative-study-failures}), common failure patterns from prior work, and through experience building and interacting with these apps. We grouped them into three main categories: UI, API, and database constraints, covering the failures in our formative study (Section~\ref{sec:formative-study-failures}) as well as more general web app expectations. We found that 29 of the 34 specific constraint cases can be checked with static methods: 21 are fully checkable, and 8 are checkable partially by structurally verifying control- and dataflow paths while deferring exact value checking to runtime. Constraints that involve exact value checking or runtime timing require heavier-weight dynamic analysis that we leave to future work.

As discussed above, every constraint consists of an action, an event, and user elements.
\textbf{Actions} can express any user action or browser event over a visible HTML element.
The user simply selects an element, and specifies the action with an optional guard.

\textbf{Events} are what occur in response to actions, and can be any write, action, API call, storage interaction, or composition of these. Table~\ref{tab:expressiveness} summarizes the main constraint patterns we encountered and express.

\begin{table*}[t]
\centering
\caption{Core expressiveness patterns of the constraint language.}
\label{tab:expressiveness}
\scriptsize
\setlength{\tabcolsep}{6pt}
\renewcommand{\arraystretch}{1.3}
\begin{tabular}{l p{7cm} l}
\textbf{Pattern} & \textbf{Description} & \textbf{Example} \\
Basic state updates & Link a user action to a basic write event. & $\tok{P(w(cartCount) | A(addBtn)) = 1}$ \\
Writes from source & Restricts a constraint to be written from a specific source. & $\tok{P(w(total, r(promoInput)) | A(applyPromoBtn)) = 1}$ \\
Compound events & Combines multiple expected events using logic operators. & $\tok{P(w(cartCount) AND w(cartItems) | A(addBtn)) = 1}$ \\
Persistence & Requires that a specific action results in saved state across page reloads. & $\tok{P(persist(cartStorage) | A(addBtn)) = 1}$ \\
API call & Requires an action to trigger an API call. & $\tok{P(call(productsAPI) | A(pageLoad)) = 1}$ \\
Disallowed events & Target event must not occur. & $\tok{P(w(cartCount) | A(applyPromoBtn)) = 0}$ \\
\end{tabular}
\end{table*}

\subsection{From Constraints to CodeQL Queries}
\label{sec:compiling-semantics}

We compile each constraint into a set of the primitive queries of Table~\ref{tab:primitives}, together with a rule for combining their pass/fail results. Figure~\ref{fig:translation-semantics} defines this compilation as five translation functions, read top-down: $\Tsem{\cdot}$ translates the whole constraint, $\Csem{\cdot}$ translates its condition side, $\Esem{h}{\cdot}$ and $\Rsem{h}{\cdot}$ translate its event side, and $\Vsem{\cdot}$ translates the value expressions that appear inside events.

\begin{table}[h]
\centering
\caption{The CodeQL primitive queries.}
\label{tab:primitives}
\scriptsize
\begin{tabular}{p{2cm}p{6cm}}
\textbf{Primitive} & \textbf{What it checks} \\
path\_exists & $\ge1$ code path from the handler writes to target.   \\
all\_paths\_write & Every exit path through the handler writes the target. \\
literal\_value & The write assigns the literal value named in the constraint. \\
source\_set & Dataflow sources to the write same as expected set. \\
self\_increment & Written value derives from $r(\text{target}) + \text{constant}$. \\
api\_result\_taint & The written value taint-flows from a fetch/axios response. \\
no\_other\_handlers & {\it Only} the named action writes to target.   \\
call\_reaches & The handler reaches an invocation of the named function. \\
call\_with\_source & r(source) flows to an argument of the call. \\
page\_load\_restores & A page-load handler calls getItem(key) for the storage key. \\
guarded\_write & The write is structurally inside an \texttt{if} reading the guarded element. \\
\end{tabular}
\end{table}

\begin{figure*}[t]
\small
\noindent\textbf{Constraint Translation}\hfill $\Tsem{\cdot}$
\begin{center}
$\Tsem{\tok{P(}E \mid C\tok{)} = 1} = \Esem{\Csem{C}}{E}
\qquad\qquad
\Tsem{\tok{P(}E \mid C\tok{)} = 0} = \neg\,\Rsem{\Csem{C}}{E}$
\end{center}
\noindent\textbf{Condition Translation}\hfill $\Csem{\cdot}$
\begin{center}
$\Csem{\tok{A(}a\tok{)}} = \handler(a)
\qquad
\Csem{\tok{A(}a\tok{)}\;\tok{AND}\;g} = \langle \handler(a),\, g\rangle
\qquad
\Csem{\tok{NOT}\;\tok{A(}a\tok{)}} = \overline{\handler(a)}$
\end{center}
\noindent\textbf{Event Translation}\hfill $\Esem{h}{\cdot}$
\begin{center}
$\begin{aligned}
\Esem{h}{\tok{w(}e\tok{)}} &= \pq{path\_exists}(h, e) \wedge \pq{all\_paths\_write}(h, e) \\
\Esem{h}{\tok{w(}e\tok{, }k\tok{)}} &= \Esem{h}{\tok{w(}e\tok{)}} \wedge \pq{literal\_value}(h, e, k) \\
\Esem{h}{\tok{w(}e\tok{, }v\tok{)}} &= \Esem{h}{\tok{w(}e\tok{)}} \wedge \pq{source\_set}(h, e, \Vsem{v}) \\
\Esem{h}{\tok{w(}e\tok{, sources=}S\tok{)}} &= \Esem{h}{\tok{w(}e\tok{)}} \wedge \pq{source\_set}(h, e, \Vsem{S}) \\
\Esem{h}{\tok{w(}e\tok{, r(}e\tok{) + }k\tok{)}} &= \Esem{h}{\tok{w(}e\tok{, r(}e\tok{))}} \wedge \pq{self\_increment}(h, e, k) \\
\Esem{h}{\tok{w(}e\tok{, r(api\_result))}} &= \Esem{h}{\tok{w(}e\tok{)}} \wedge \pq{api\_result\_taint}(h, e) \\
\Esem{h}{\tok{call(}c\tok{)}} &= \pq{call\_reaches}(h, c) \\
\Esem{h}{\tok{call(}c\tok{, }v\tok{)}} &= \pq{call\_with\_source}(h, c, \Vsem{v}) \\
\Esem{h}{\tok{persist(}s\tok{)}} &= \pq{path\_exists}(h, s) \wedge \pq{page\_load\_restores}(s)
\end{aligned}$

\medskip
$\begin{aligned}
\Esem{h}{E_1\;\tok{AND}\;E_2} &= \Esem{h}{E_1} \wedge \Esem{h}{E_2} \\
\Esem{h}{E_1\;\tok{XOR}\;E_2} &= \Esem{h}{E_1} \oplus \Esem{h}{E_2}
\end{aligned}
\qquad\qquad
\begin{aligned}
\Esem{h}{E_1\;\tok{OR}\;E_2} &= \Esem{h}{E_1} \vee \Esem{h}{E_2} \\
\Esem{h}{\tok{NOT}\;E} &= \neg\,\Esem{h}{E}
\end{aligned}$

\medskip
$\Esem{\langle h,\, g\rangle}{\tok{w(}e\tok{)}} = \Esem{h}{\tok{w(}e\tok{)}} \wedge \pq{guarded\_write}(h, e, g)
\qquad
\Esem{\overline{\handler(a)}}{\tok{w(}e\tok{)}} = \pq{no\_other\_handlers}(a, e)$
\end{center}
\noindent\textbf{Reachability Translation (for $P = 0$)}\hfill $\Rsem{h}{\cdot}$
\begin{center}
$\begin{aligned}
\Rsem{h}{\tok{w(}e\tok{, \ldots)}} &= \pq{path\_exists}(h, e) \\
\Rsem{h}{\tok{call(}c\tok{, \ldots)}} &= \pq{call\_reaches}(h, c) \\
\Rsem{h}{\tok{persist(}s\tok{)}} &= \pq{path\_exists}(h, s)
\end{aligned}
\qquad\qquad
\begin{aligned}
\Rsem{h}{E_1 \mathbin{op} E_2} &= \Rsem{h}{E_1} \mathbin{\widehat{op}} \Rsem{h}{E_2} \\
\Rsem{h}{\tok{NOT}\;E} &= \neg\,\Rsem{h}{E}
\end{aligned}$
\end{center}
\noindent\textbf{Value Translation (expected sources)}\hfill $\Vsem{\cdot}$
\begin{center}
$\Vsem{\tok{r(}x\tok{)}} = \{x\}
\qquad
\Vsem{k} = \emptyset
\qquad
\Vsem{v_1 \mathbin{op} v_2} = \Vsem{v_1} \cup \Vsem{v_2}
\qquad
\Vsem{\tok{\{}v_1\tok{, }\ldots\tok{, }v_n\tok{\}}} = \Vsem{v_1} \cup \cdots \cup \Vsem{v_n}$
\end{center}
\caption{The translation semantics from constraints to CodeQL primitive queries. Primitives are the queries of Table~\ref{tab:primitives}, run with handler context $h$ as their entry point; the auxiliary function $\handler(a)$ resolves element id $a$ to the event-handler function the app registers on it. Metavariables: $a$ ranges over action elements, $e$ over element ids, $s$ over storage keys, $c$ over APIs, $k$ over literals, $v$ over value expressions, $S$ over source sets, $g$ over guards, and $E$, $C$ over event and condition expressions. When several $\mathsf{E}$ rules overlap, the most specific applies. In $\mathsf{V}$, $op \in \{\tok{+}, \tok{-}, \tok{*}, \tok{/}\}$; in $\mathsf{R}$, $op \in \{\tok{AND}, \tok{OR}, \tok{XOR}\}$ and $\widehat{op}$ is the corresponding Boolean connective ($\wedge$, $\vee$, $\oplus$).}
\label{fig:translation-semantics}
\end{figure*}

\smallskip \noindent \textbf{Constraints ($\Tsem{\cdot}$).}
The probability decides how the event side is checked. A $P{=}1$ constraint requires the event on \emph{every} path, so its event is translated by $\Esem{h}{\cdot}$, whose primitives check all paths. A $P{=}0$ constraint forbids the event on \emph{any} path, so its event is translated by $\Rsem{h}{\cdot}$, which only asks whether the event is reachable at all; the constraint passes when it is not. For persist, this means the action handler never writes the storage key.

\smallskip \noindent \textbf{Conditions ($\Csem{\cdot}$).}
The condition side is translated once into the \emph{handler context} $h$ from which every primitive query begins. In the common case, $h$ is simply $\handler(a)$. The function $\handler(\cdot)$ is not part of the constraint language, but an auxiliary function used by the translation: given the id $a$ of an interface element, it returns the event-handler function that the app registers on that element.

$\handler(a)$ resolves the id to its code references, including calls such as \texttt{getElementById("addBtn")} and the variables to which they are assigned (e.g., \texttt{isElementRef} predicate in the Appendix). We then identify the function attached to the element's events. Each query on the event side searches only code reachable from that function.

A guard keeps the handler but records the guard, written $\langle\handler(a), g\rangle$; under this context a write must additionally sit inside a conditional that reads the guarded element (\pq{guarded\_write}). A negated action instead changes the context to $\overline{\handler(a)}$, and writes are checked using \pq{no\_other\_handlers} so that only $a$ can write the target.

The distinction between the two forms of negation appears directly in the rules. The constraint $P(w(c) \mid A(a)) = 0$ asks whether $a$'s own handler can reach the write and passes when it cannot. 
By contrast, $P(w(c) \mid \tok{NOT}\;A(a)) = 1$ checks exclusivity, passing when no \emph{other} handler reaches the write.

\smallskip \noindent \textbf{Events ($\Esem{h}{\cdot}$) and values ($\Vsem{\cdot}$).}
$\Esem{h}{\cdot}$ follows the structure of the event expression and covers the three checkable event atoms: writes, calls, and persists. User actions and guards instead belong to the condition side and are handled by $\Csem{\cdot}$. 
A plain write $\tok{w(}e\tok{)}$ produces two primitives: \pq{path\_exists}, requiring the write to be reachable from $h$, and \pq{all\_paths\_write}, requiring the write to occur on every path through the handler. A write includes either an assignment to the element (e.g., \texttt{.textContent}, \texttt{.value}) or a DOM mutation on it. Additional details in the atom add primitives to this base---for instance, a literal value adds \pq{literal\_value}, a value expression adds \pq{source\_set}.

A \tok{call} atom checks that the handler reaches the named API. When the atom names a source value, \pq{call\_with\_source} checks both that the call is reachable and that the source flows into it. A \tok{persist} atom checks both parts of persistence: the handler must save to the storage key using \pq{path\_exists}, and a page-load handler must read the value back using \pq{page\_load\_restores}. Compound events are translated one operand at a time, after which their pass/fail results are combined using the Boolean operator in the constraint: \tok{AND} as $\wedge$, \tok{OR} as $\vee$, \tok{XOR} as $\oplus$, and \tok{NOT} as $\neg$.

Each primitive corresponds to one CodeQL query executed from entry point $h$. The query returns a set of rows, with each row representing one match in the code, and the primitive interprets those rows as pass/fail. For example, \pq{path\_exists} passes when the query finds at least one reachable write. By contrast, \pq{all\_paths\_write} searches for counterexample paths that exit without writing, and passes only when none exist. Finally, $\Vsem{\cdot}$ does not evaluate arithmetic. It only collects the components read by a value expression, so $\tok{r(}a\tok{)} + \tok{r(}b\tok{)}$ becomes the source set \${a,b}\$. The checker verifies where a value comes from, rather than what the expression computes.

\begin{example}
Let $h = \handler(\tok{applyPromoBtn})$. The promo-code constraint from the introduction compiles by the rules of Figure~\ref{fig:translation-semantics} as:
{\small\begin{align*}
&\mathsf{T}\llbracket\tok{P(w(total, r(promoInput))} 
\tok{| A(applyPromoBtn)) = 1}\rrbracket \\
&\quad= \Esem{h}{\tok{w(total, r(promoInput))}} \\
&\quad= \pq{path\_exists}(h, \tok{total}) \\
&\qquad\; \wedge\ \pq{all\_paths\_write}(h, \tok{total}) \\
&\qquad\; \wedge\ \pq{source\_set}(h, \tok{total}, \{\tok{promoInput}\})
\end{align*}}
In the broken app, the handler writes only the success message, so \pq{all\_paths\_write} returns counterexample paths that never write \tok{total}, and the constraint fails.
\end{example}

\subsection{Why We Chose This Design}
Other formalisms like Hoare logic or Linear Temporal Logic can be more expressive, but are less accessible to non-developers. Writing our promo example in Hoare logic as $\{P\}$ applyPromo() $\{Q\}$, requires the user to write out the DOM state before and after, as well as read the code for internal function names and variables. A non-developer, or vibe-coder, would not know these, and learning this would take as long as debugging manually. By restricting constraints to user-visible behavior, users express their expectations easily over their own app.

\section{Implementation}
\label{sec:implementation}
To go from the app to verified constraints, we follow three steps: (1) preprocess the code to add identifiers (Section~\ref{sec:preprocessing}), (2) run the app for authoring constraints (Section~\ref{sec:ui}), and (3) parse, validate, and verify the constraint (Section~\ref{sec:analysis}).

\subsection{Preprocessing}
\label{sec:preprocessing}
Vibe-coded apps often lack identifiers for elements~\cite{https://doi.org/10.1002/smr.1771}. This creates two issues: (1) when authoring a constraint, we have no id to uniquely reference each element; (2) the id links the HTML and JS, which both reference the same DOM elements. To address this, we automatically map elements across both languages by scanning the source code to locate all UI elements (e.g. buttons and inputs), and adding unique identifiers (\texttt{cv\_nnnn}) where they are missing. This ensures every element has a static reference across both languages. We use BeautifulSoup to parse each HTML file and traverse the elements the user could select. For JS, we use regular expressions to identify programmatically created elements (such as \texttt{createElement()}), and append missing IDs.

\subsection{User Interface}
\label{sec:ui}
Users provide \systemname with their web app's path, which opens in a new tab. Users express constraints on this app via our overlay template: ``When I take [action], these update: [component]'' with optional AND, OR, XOR, NOT logic.

Users select visible UI components by clicking directly on them, the same way they interact with their app. They can also pick from a detected list of APIs and storage. As a result, the user does not need to learn the language syntax, identify internal variables for elements, or write constraints by hand. \systemname records the user's actions, converts them to our constraint language, and sends them to the parser. 

\begin{figure}[h]
    \centering
\includegraphics[width=\columnwidth]{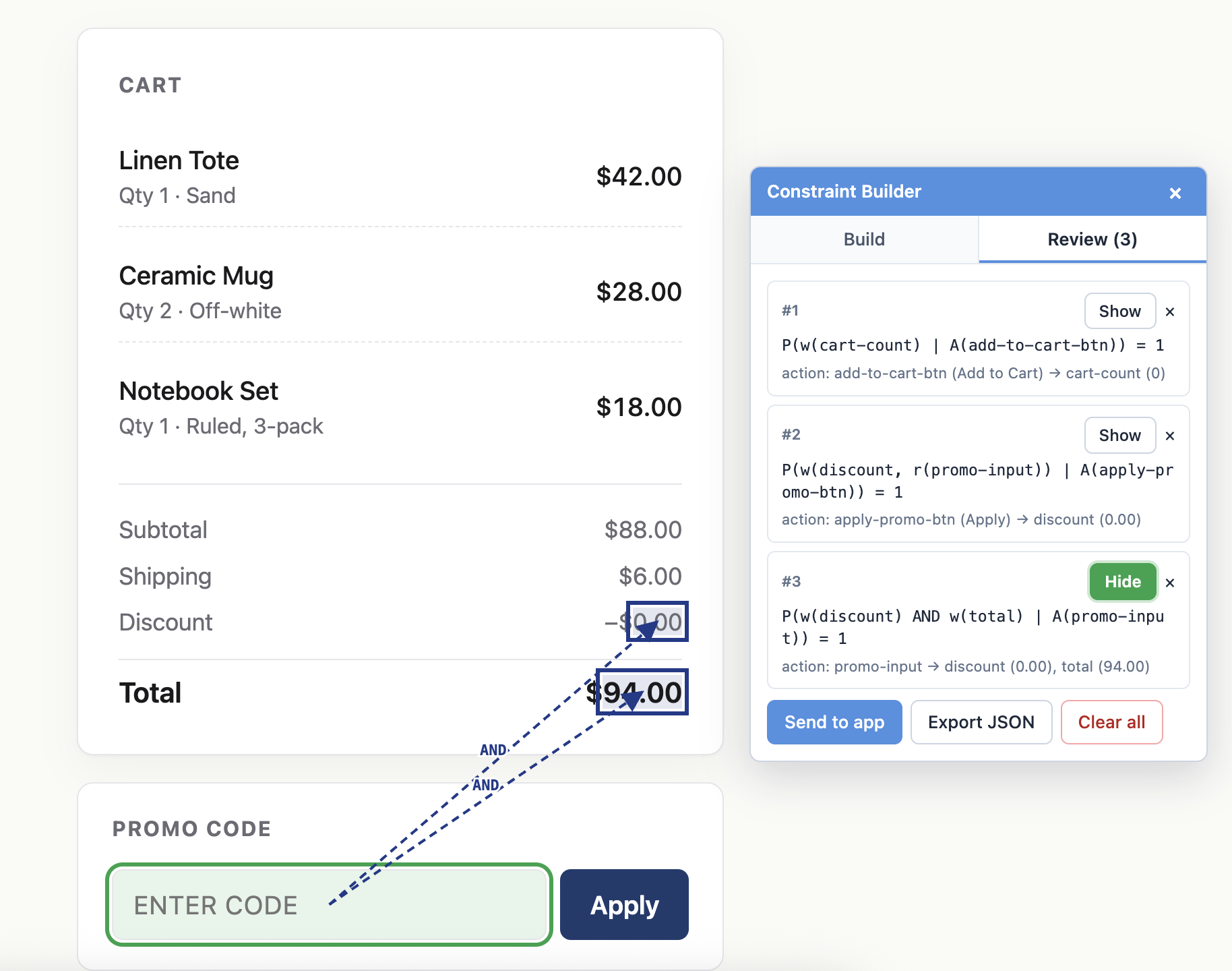}
    \caption{Reviewing authored constraints. The Review panel lists the constraints the user has built, and dashed lines trace each constraint's flows from action to event.}
    \label{fig:constraint-visual}
    \Description{Screenshot of the same cart page with the Review tab of the constraint builder open. Three authored constraints are listed in the panel, and dashed arrows connect the promo input and Apply button on the page to the discount and total fields they write to.}
\end{figure}

\subsection{Analysis}
\label{sec:analysis}
Once the user selects their components through the UI, their selections are translated into our constraint language, and compiled to static CodeQL queries in 4 steps:

\begin{enumerate}[nosep, leftmargin=1.2em]
\item \textbf{Parsing:} Constraints are parsed using an ANTLR4-generated parser into an AST, using our grammar from Section ~\ref{sec:language}.
    
\item \textbf{Semantic analysis:} We walk the AST to extract the key parts (the event, condition, probability, and identifiers) and verify semantic rules: every referenced identifier must match a valid DOM element, storage identifier, or API, the condition side must contain an action $A(e_i)$, the event side must contain a valid effect ($w()$, $call()$, $persist()$), the probability must be 0 or 1, and identifiers in action must map to the ``action'' type. If rule fails, the constraint is invalid and exits with a detailed message to the user. 
    
\item \textbf{Classification}:
    From this AST, we determine which primitive queries to run based on the event type and condition (Section \ref{sec:compiling-semantics}). For example, an event containing an API node resolves to an API call check, while a write carrying a value expression resolves to a dataflow check.
    
\item \textbf{Verification}:
    Once we know which primitives to run, \systemname executes each corresponding CodeQL query--written in a \texttt{.ql} file with placeholder tokens replaced by element identifiers--against the database (Section \ref{sec:overview}). It reviews the returned rows to return a Pass/Fail result, and reports which queries failed for the user to review.
\end{enumerate}

\subsection{Authoring Constraints}
\label{sec:authoring}
Constraints can be generated in 2 ways:
\begin{enumerate} [nosep, leftmargin=*]
    \item \textbf{User:} The user can author constraints through our overlay UI, which lets a user click on elements in their running app and pick from constraint templates. For this, they do not need to know the constraint syntax or understand the code, only simple logic expressions (AND/OR/XOR/NOT).

    \item \textbf{LLM or Agent:} We describe the language as a skill for an agent or LLM, containing the grammar, constraint types, and examples. Given the code, the app description, and this context, a model can produce a comprehensive list of constraints, which can combine with the user's set for broader coverage. In Section~\ref{sec:constraint-probe} we find that models can author valid constraints and use them to catch more bugs.

\end{enumerate}
\section{Evaluation}
In our evaluation, we focused on two questions: (1) does \systemname catch the constraint violations it is designed to catch, and (2) how does it compare to asking a frontier LLM to find the same bugs?

\subsection{Experiment Setup}

\subsubsection{Test Applications}
Test applications were generated using Claude Code. We built four web apps, each modeled after an existing app (Amazon, Twitter, Airbnb, Slack). We chose well-known apps because they are real-world use cases, and models understand the expected features and behavior due to extensive training data. We used JS with no frameworks to stay within scope of static analysis, ensured all elements had IDs, and used local storage. We iterated on each app, adding features to create opportunities to write constraints. For example, a local database of items for Amazon, or a fees calculator for Airbnb.

\subsubsection{Ground Truth and Injected Violations} 
We established 14--21 constraints per app representing the expected behavior (e.g. applying a promo code updates the total). We copied each app to \texttt{modified/} variant and guided an agent to introduce 7--8 subtle violations, breaking about half the constraints, to simulate real-world development mistakes. These were inspired by documented coding-agent failure patterns~\cite{daplab9patterns}, taxonomies of LLM-generated bugs~\cite{10.1007/s10664-025-10614-4}, our formative study (Section~\ref{sec:formative-study-failures}), and by manually breaking or disconnecting components. Bugs covered several types, including missing branch write (if x then \{ write \} with no else), switch missing a default case, write-with-wrong-source, and disconnected components. For example, in the modified Twitter app, post-tweet-btn writes the posted-banner even for empty inputs, dropping the character count check. The apps ranged from roughly 300 to 900 lines of code (HTML and JS), each with 16–26 action elements, 17–23 event handlers, and 5–11 localStorage operations. This added 30 behavioral violations across the four applications. We evaluated whether \systemname could correctly flag these constraint violations, and compared it against the LLM baselines.

\subsubsection{Methods}
We compared \systemname against asking a frontier LLM to find bugs in the same broken code. We tested three models: Claude Opus 4.7, DeepSeek V3, and Gemini Pro, using three prompt variants, increasing in detail (using a single run per model-prompt cell).

\stitle{Prompt 1:} ``Here is a web app, similar to [well known app]. Are there any bugs?''

\stitle{Prompt 2:} Lists the features the user requested, e.g. ``the user requested an Amazon like app with these features: a product grid, a cart drawer\ldots'' 

\stitle{Prompt 3:} Same feature list as Prompt 2 plus an explicit edge-case checklist: boundary values, all user states, all branches, cross-handler consistency.

\subsection{Results}
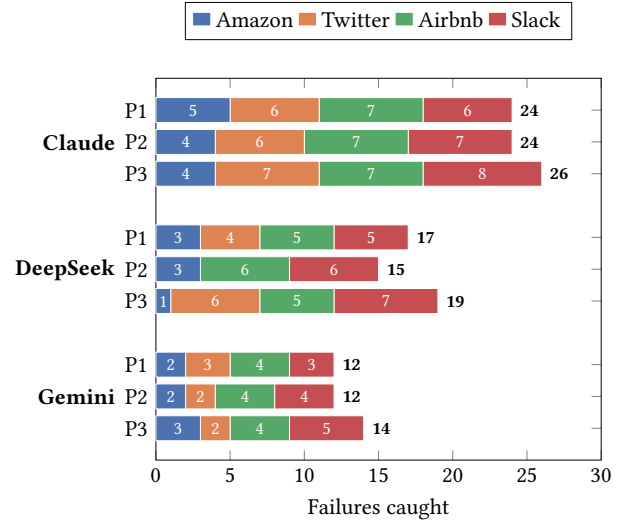
\begin{figure}[tb]
\centering
\resizebox{\linewidth}{!}{%
\definecolor{cAmazon}{HTML}{4C72B0}
\definecolor{cTwitter}{HTML}{DD8452}
\definecolor{cAirbnb}{HTML}{55A868}
\definecolor{cSlack}{HTML}{C44E52}
\begin{tikzpicture}
\begin{axis}[
  xbar stacked,
  y dir=reverse,
  xmin=0, xmax=30,
  width=\linewidth, height=6.5cm,
  bar width=11pt,
  y=14pt,
  ytick=data,
  yticklabels={P1,P2,P3,P1,P2,P3,P1,P2,P3},
  xlabel={Failures caught},
  every node near coord/.append style={
    white, font=\footnotesize\bfseries, anchor=center
  },
  nodes near coords,
  legend style={at={(0.5,1.1)},anchor=south,legend columns=4},
  clip=false,
]
\addplot+[fill=cAmazon, draw=white] coordinates {(5,0)(4,1)(4,2) (3,4)(3,5)(1,6) (2,8)(2,9)(3,10)};
\addplot+[fill=cTwitter, draw=white] coordinates {(6,0)(6,1)(7,2) (4,4)(0,5)(6,6) (3,8)(2,9)(2,10)};
\addplot+[fill=cAirbnb, draw=white] coordinates {(7,0)(7,1)(7,2) (5,4)(6,5)(5,6) (4,8)(4,9)(4,10)};
\addplot+[fill=cSlack, draw=white] coordinates {(6,0)(7,1)(8,2) (5,4)(6,5)(7,6) (3,8)(4,9)(5,10)};
\legend{Amazon,Twitter,Airbnb,Slack}
\node[anchor=west, font=\footnotesize\bfseries] at (axis cs:24,0) {24};
\node[anchor=west, font=\footnotesize\bfseries] at (axis cs:24,1) {24};
\node[anchor=west, font=\footnotesize\bfseries] at (axis cs:26,2) {26};
\node[anchor=west, font=\footnotesize\bfseries] at (axis cs:17,4) {17};
\node[anchor=west, font=\footnotesize\bfseries] at (axis cs:15,5) {15};
\node[anchor=west, font=\footnotesize\bfseries] at (axis cs:19,6) {19};
\node[anchor=west, font=\footnotesize\bfseries] at (axis cs:12,8) {12};
\node[anchor=west, font=\footnotesize\bfseries] at (axis cs:12,9) {12};
\node[anchor=west, font=\footnotesize\bfseries] at (axis cs:14,10){14};
\node[anchor=east, font=\bfseries] at (axis cs:-2.2,1)  {Claude};
\node[anchor=east, font=\bfseries] at (axis cs:-2.2,5)  {DeepSeek};
\node[anchor=east, font=\bfseries] at (axis cs:-2.2,9)  {Gemini};
\end{axis}
\end{tikzpicture}
}
\caption{Bug catch rate of \systemname compared to the LLM baselines (Claude, DeepSeek, Gemini) across the three prompt variants. \systemname flags all injected violations, while the best model--prompt pair reaches 26/30.}
\label{fig:catch-rates}
\end{figure}

\systemname catches all constraint violations, while LLMs were slower and less accurate, with the best model-prompt pair (Claude P3) reaching 26/30 and most results far lower (Figure~\ref{fig:catch-rates}). The results suggest that models are better at spotting localized bugs, a wrong literal or a missing write inside one handler, but struggle once bugs span multiple branches or handlers. Providing more detail in the prompt does not necessarily help; the models may read more of the code but not always find more of the bugs.

\subsubsection{What Models Catch}
Models reliably caught bugs localized to a small code block, whose patterns are relatively easy to recognize. For example, assigning an incorrect literal, or failing to update the UI on one conditional branch. Models also caught writes from the wrong data source (e.g. apply-promo saving the old value instead of the user's input) and partial persistence (e.g. storing favorites but not loading them on refresh). This was common in the Airbnb app, whose bugs largely involve localized writes that were missing or wrote the wrong key. Identifying these bugs does not require any control flow analysis or handler interaction.

\subsubsection{What Models Miss}
We found two dominant classes of violations that LLMs did not reliably identify.
The first are universal quantifier constraints, where a property must hold over all paths, such as P=1 writes. Verifying these requires checking all branches, but models sometimes miss the branch where the write is dropped. For example, an Airbnb booking handler that updates the total cost only when guests $\leq$ 4, but leaves the other branch silently broken.
The second are cross-handler flows, where one handler's correctness depends on state written by another. These are difficult to detect because the bug is not visible in either handler alone, only when they interact. For example, Amazon’s checkout reads \texttt{cartSummary} but \texttt{applyPromo} clears it to 0. Since the handlers never reference each other, models reason on them separately and don’t trace the data flow between them.

We found that adding detail to the prompt does not always help; all three prompt versions did not reliably catch errors across models. With more detail, models read code more thoroughly, but grew more willing to trust it, justifying bugs instead. For example, Amazon's laptop favorite handler silently skips its write for one user tier, but a model dismissed this as intentional:
\emph{"USER\_TIER is a const initialized to 'free', so the enterprise branch is dead code, not a defect."}

\subsubsection{Agents and our Constraint Language}
\label{sec:constraint-probe}
We used the Amazon app as a probe to understand whether LLMs can benefit from \systemname. We studied two questions: (Q1) can agents author constraints? (Q2) does providing our language and asking the model to write constraints help find more bugs? To do so, we performed three runs; in each, we prompted each model with the grammar, a description of the operators, example constraints, a description of the Amazon app, and the Amazon code, and asked it to generate a comprehensive set of constraints to find bugs. We consider a constraint valid if it parses and passes our semantic rules.

\smallskip\stitle{Q1.} Per run, Claude, DeepSeek, and Gemini each generated 30-60, 50-86, and 15-20 constraints.  Of these, 45-100\% were valid for Gemini, 34-44\% for Claude, and 41-65\% for DeepSeek.  Although they nearly all made sense conceptually, their errors were finer: hallucinated names (e.g. \texttt{dbPut\_favorites}), syntax errors (underscore instead of dash), or the wrong atom for an event (\texttt{call(x)} instead of \texttt{A(x)}). Since they expressed reasonable expectations and failed on syntax or identifiers, this could likely be improved via more prompt engineering.

\smallskip
\stitle{Q2.} The constraints helped models catch more of the silent failures: compared to using their best prompt, Claude increased from $5\to 6$, DeepSeek from $3\to5$, and Gemini from $3\to4$ out of 7 gold bugs. DeepSeek likely improved the most because it generated the most constraints. Among the bugs the models identified, Claude and Gemini produced no false positives, while DeepSeek produced 2-3 per run. Claude and DeepSeek also identified other bugs out of scope of \systemname. While this is not conclusive evidence, it is a promising signal that providing a compact constraint language can help improve bug catch rates.

\section{Limitations and Future Work}
\subsection{Limitations}
CodeQL is a static analyzer that reads the code without running, so it cannot link JS to DOM elements created at runtime. For example, elements built through strings (\texttt{container.\allowbreak innerHTML="..."}) hide the ids inside the string, since the value is never parsed further. As a workaround, users can write constraints at the static container level. Frameworks like React and Vue have the same issue since they build  UIs at runtime from a virtual DOM. This usually leaves the static HTML empty and \systemname cannot fully check them. Drag-and-drop is similar: implementations vary across the HTML5 drag API, touch events, and mouse movement, with no single code pattern, making it difficult to detect statically.

\subsection{Future Work}
\textbf{Runtime Analysis:}
Our language supports probability values within [0, 1], but our implementation focuses on P=1 and P=0. Handling exact values and timing bugs (e.g., race conditions, API status codes) requires runtime data. We plan to address this using a browser agent to collect runtime traces, verify constraints, and record the interactions that led to the error to help debugging.

Runtime tracing introduces variance because results depend on browser agent paths, raising open questions about action selection and modeling users, which is why we focused on static analysis first. A promising direction is a hybrid approach: using static analysis for structural checks, and using runtime tracing only for value or timing related checks. For example, the constraint $P(w(e_j, \text{len}(r(\text{api\_result}))) \mid \text{call}(\text{api})) = 1$ expresses that a component ej must display the length of an API result. The static checker confirms the value written to ej derives from the API response length, and runtime traces verify the resulting length matches.

\textbf{User Evaluation:} While our evaluation demonstrates that \systemname can accurately detect injected failures, we plan to conduct a user study with vibe-coders to evaluate language usability, how users interpret analysis results, and how they integrate them into their coding workflow.

\textbf{Iterative Vibe Coding:} \systemname evaluates constraints on single code versions, but vibe coding is an iterative process where constraints change as apps grow. A key next step is tracking and evolving constraints alongside code versions, allowing coding agents to continuously generate, check, and refine constraints and code throughout development.
\section{Conclusion}
We introduced a constraint language enabling users to express expected web-app behavior simply by pointing at the app itself. We also introduced \systemname, a tool that verifies these constraints against their code through static analysis. We do so by parsing the constraint, classifying the type, and mapping to a set of CodeQL queries that we then execute. In our evaluation, \systemname caught all 30/30 injected violations, compared to at most 26/30 for the strongest LLM baseline and fewer for the rest. Combined with the authoring interface, this lets users of any programming background express and verify expectations deterministically, without needing to read or understand the code.

\bibliographystyle{ACM-Reference-Format}

\bibliography{references}

@misc{dente2026constraintdecayfragilityllm,
      title={Constraint Decay: The Fragility of LLM Agents in Backend Code Generation}, 
      author={Francesco Dente and Dario Satriani and Paolo Papotti},
      year={2026},
      eprint={2605.06445},
      archivePrefix={arXiv},
      primaryClass={cs.SE},
      url={https://arxiv.org/abs/2605.06445}, 
}

@inproceedings{10.1145/3786335.3813180,
author = {Tran, Hung and Nashold, Langston and Krishnan, Rayan and Bigeard, Antoine and Gu, Alex},
title = {Vibe Code Bench: Evaluating AI Models on End-to-End Web Application Development},
year = {2026},
isbn = {9798400724152},
publisher = {Association for Computing Machinery},
address = {New York, NY, USA},
url = {https://doi.org/10.1145/3786335.3813180},
doi = {10.1145/3786335.3813180},
booktitle = {Proceedings of the ACM Conference on AI and Agentic Systems},
pages = {514–536},
numpages = {23},
location = {
},
series = {CAIS '26}
}

@misc{vero2025baxbenchllmsgeneratecorrect,
      title={BaxBench: Can LLMs Generate Correct and Secure Backends?}, 
      author={Mark Vero and Niels Mündler and Victor Chibotaru and Veselin Raychev and Maximilian Baader and Nikola Jovanović and Jingxuan He and Martin Vechev},
      year={2025},
      eprint={2502.11844},
      archivePrefix={arXiv},
      primaryClass={cs.CR},
      url={https://arxiv.org/abs/2502.11844}, 
}

@inproceedings {246326,
author = {Chang Lou and Peng Huang and Scott Smith},
title = {Understanding, Detecting and Localizing Partial Failures in Large System Software },
booktitle = {17th USENIX Symposium on Networked Systems Design and Implementation (NSDI 20)},
year = {2020},
isbn = {978-1-939133-13-7},
address = {Santa Clara, CA},
pages = {559--574},
url = {https://www.usenix.org/conference/nsdi20/presentation/lou},
publisher = {USENIX Association},
month = feb
}

@misc{jain2024livecodebenchholisticcontaminationfree,
      title={LiveCodeBench: Holistic and Contamination Free Evaluation of Large Language Models for Code}, 
      author={Naman Jain and King Han and Alex Gu and Wen-Ding Li and Fanjia Yan and Tianjun Zhang and Sida Wang and Armando Solar-Lezama and Koushik Sen and Ion Stoica},
      year={2024},
      eprint={2403.07974},
      archivePrefix={arXiv},
      primaryClass={cs.SE},
      url={https://arxiv.org/abs/2403.07974}, 
}

@inproceedings {280920,
author = {Chang Lou and Yuzhuo Jing and Peng Huang},
title = {Demystifying and Checking Silent Semantic Violations in Large Distributed Systems},
booktitle = {16th USENIX Symposium on Operating Systems Design and Implementation (OSDI 22)},
year = {2022},
isbn = {978-1-939133-28-1},
address = {Carlsbad, CA},
pages = {91--107},
url = {https://www.usenix.org/conference/osdi22/presentation/lou-demystifying},
publisher = {USENIX Association},
month = jul
}

@misc{tian2024debugbenchevaluatingdebuggingcapability,
      title={DebugBench: Evaluating Debugging Capability of Large Language Models}, 
      author={Runchu Tian and Yining Ye and Yujia Qin and Xin Cong and Yankai Lin and Yinxu Pan and Yesai Wu and Haotian Hui and Weichuan Liu and Zhiyuan Liu and Maosong Sun},
      year={2024},
      eprint={2401.04621},
      archivePrefix={arXiv},
      primaryClass={cs.SE},
      url={https://arxiv.org/abs/2401.04621}, 
}

@misc{jimenez2024swebenchlanguagemodelsresolve,
      title={SWE-bench: Can Language Models Resolve Real-World GitHub Issues?}, 
      author={Carlos E. Jimenez and John Yang and Alexander Wettig and Shunyu Yao and Kexin Pei and Ofir Press and Karthik Narasimhan},
      year={2024},
      eprint={2310.06770},
      archivePrefix={arXiv},
      primaryClass={cs.CL},
      url={https://arxiv.org/abs/2310.06770}, 
}

@inproceedings{10.1145/3759425.3763390,
author = {Mitchell, Jacqueline and Shaaban, Yasser},
title = {Position: Vibe Coding Needs Vibe Reasoning: Improving Vibe Coding with Formal Verification},
year = {2025},
isbn = {9798400721489},
publisher = {Association for Computing Machinery},
address = {New York, NY, USA},
url = {https://doi.org/10.1145/3759425.3763390},
doi = {10.1145/3759425.3763390},
booktitle = {Proceedings of the 1st ACM SIGPLAN International Workshop on Language Models and Programming Languages},
pages = {84–90},
numpages = {7},
location = {Singapore, Singapore},
series = {LMPL '25}
}

@misc{ma2025specgenautomatedgenerationformal,
      title={SpecGen: Automated Generation of Formal Program Specifications via Large Language Models}, 
      author={Lezhi Ma and Shangqing Liu and Yi Li and Xiaofei Xie and Lei Bu},
      year={2025},
      eprint={2401.08807},
      archivePrefix={arXiv},
      primaryClass={cs.SE},
      url={https://arxiv.org/abs/2401.08807}, 
}

@misc{sun2024cloverclosedloopverifiablecode,
      title={Clover: Closed-Loop Verifiable Code Generation}, 
      author={Chuyue Sun and Ying Sheng and Oded Padon and Clark Barrett},
      year={2024},
      eprint={2310.17807},
      archivePrefix={arXiv},
      primaryClass={cs.AI},
      url={https://arxiv.org/abs/2310.17807}, 
}

@misc{zuo2025patagentautoformalizationmodelchecking,
      title={PAT-Agent: Autoformalization for Model Checking}, 
      author={Xinyue Zuo and Yifan Zhang and Hongshu Wang and Yufan Cai and Zhe Hou and Jing Sun and Jin Song Dong},
      year={2025},
      eprint={2509.23675},
      archivePrefix={arXiv},
      primaryClass={cs.SE},
      url={https://arxiv.org/abs/2509.23675}, 
}

@inproceedings{10.1145/3411764.3445567,
author = {Zhao, Valerie and Zhang, Lefan and Wang, Bo and Littman, Michael L. and Lu, Shan and Ur, Blase},
title = {Understanding Trigger-Action Programs Through Novel Visualizations of Program Differences},
year = {2021},
isbn = {9781450380966},
publisher = {Association for Computing Machinery},
address = {New York, NY, USA},
url = {https://doi.org/10.1145/3411764.3445567},
doi = {10.1145/3411764.3445567},
booktitle = {Proceedings of the 2021 CHI Conference on Human Factors in Computing Systems},
articleno = {312},
numpages = {17},
location = {Yokohama, Japan},
series = {CHI '21}
}

@inproceedings{10.1145/985692.985712,
author = {Ko, Amy J. and Myers, Brad A.},
title = {Designing the whyline: a debugging interface for asking questions about program behavior},
year = {2004},
isbn = {1581137028},
publisher = {Association for Computing Machinery},
address = {New York, NY, USA},
url = {https://doi.org/10.1145/985692.985712},
doi = {10.1145/985692.985712},
booktitle = {Proceedings of the SIGCHI Conference on Human Factors in Computing Systems},
pages = {151–158},
numpages = {8},
location = {Vienna, Austria},
series = {CHI '04}
}

@INPROCEEDINGS{1201191,
  author={Burnett, M. and Cook, C. and Pendse, O. and Rothermel, G. and Summet, J. and Wallace, C.},
  booktitle={25th International Conference on Software Engineering, 2003. Proceedings.}, 
  title={End-user software engineering with assertions in the spreadsheet paradigm}, 
  year={2003},
  volume={},
  number={},
  pages={93-103},
  doi={10.1109/ICSE.2003.1201191}}

@misc{bansal2026vibepassvibecodersreally,
      title={VIBEPASS: Can Vibe Coders Really Pass the Vibe Check?}, 
      author={Srijan Bansal and Jiao Fangkai and Yilun Zhou and Austin Xu and Shafiq Joty and Semih Yavuz},
      year={2026},
      eprint={2603.15921},
      archivePrefix={arXiv},
      primaryClass={cs.SE},
      url={https://arxiv.org/abs/2603.15921}, 
}

@misc{wang2025codesemanticshelpcomprehensive,
      title={Do Code Semantics Help? A Comprehensive Study on Execution Trace-Based Information for Code Large Language Models}, 
      author={Jian Wang and Xiaofei Xie and Qiang Hu and Shangqing Liu and Yi Li},
      year={2025},
      eprint={2509.11686},
      archivePrefix={arXiv},
      primaryClass={cs.SE},
      url={https://arxiv.org/abs/2509.11686}, 
}

@misc{chen2023teachinglargelanguagemodels,
      title={Teaching Large Language Models to Self-Debug}, 
      author={Xinyun Chen and Maxwell Lin and Nathanael Schärli and Denny Zhou},
      year={2023},
      eprint={2304.05128},
      archivePrefix={arXiv},
      primaryClass={cs.CL},
      url={https://arxiv.org/abs/2304.05128}, 
}

@article{https://doi.org/10.1002/smr.1771,
author = {Leotta, Maurizio and Stocco, Andrea and Ricca, Filippo and Tonella, Paolo},
title = {Robula+: an algorithm for generating robust XPath locators for web testing},
journal = {Journal of Software: Evolution and Process},
volume = {28},
number = {3},
pages = {177-204},
doi = {https://doi.org/10.1002/smr.1771},
url = {https://onlinelibrary.wiley.com/doi/abs/10.1002/smr.1771},
eprint = {https://onlinelibrary.wiley.com/doi/pdf/10.1002/smr.1771},
year = {2016}
}

@article{10.1007/s10515-013-0128-9,
author = {Nguyen, Bao N. and Robbins, Bryan and Banerjee, Ishan and Memon, Atif},
title = {GUITAR: an innovative tool for automated testing of GUI-driven software},
year = {2014},
issue_date = {March     2014},
publisher = {Kluwer Academic Publishers},
address = {USA},
volume = {21},
number = {1},
issn = {0928-8910},
url = {https://doi.org/10.1007/s10515-013-0128-9},
doi = {10.1007/s10515-013-0128-9},
journal = {Automated Software Engg.},
month = mar,
pages = {65–105},
numpages = {41}
}

@article{10.1145/2109205.2109208,
author = {Mesbah, Ali and van Deursen, Arie and Lenselink, Stefan},
title = {Crawling Ajax-Based Web Applications through Dynamic Analysis of User Interface State Changes},
year = {2012},
issue_date = {March 2012},
publisher = {Association for Computing Machinery},
address = {New York, NY, USA},
volume = {6},
number = {1},
issn = {1559-1131},
url = {https://doi.org/10.1145/2109205.2109208},
doi = {10.1145/2109205.2109208},
journal = {ACM Trans. Web},
month = mar,
articleno = {3},
numpages = {30}
}

@ARTICLE{5728834,
  author={Mesbah, Ali and van Deursen, Arie and Roest, Danny},
  journal={IEEE Transactions on Software Engineering}, 
  title={Invariant-Based Automatic Testing of Modern Web Applications}, 
  year={2012},
  volume={38},
  number={1},
  pages={35-53},
  doi={10.1109/TSE.2011.28}}

@inproceedings{302163.302183,
author = {Rothermel, Gregg and Li, Lixin and DuPuis, Christopher and Burnett, Margaret},
title = {What you see is what you test: a methodology for testing form-based visual programs},
year = {1998},
isbn = {0818683686},
publisher = {IEEE Computer Society},
address = {USA},
booktitle = {Proceedings of the 20th International Conference on Software Engineering},
pages = {198–207},
numpages = {10},
location = {Kyoto, Japan},
series = {ICSE '98}
}

@misc{wen2024enchantingprogramspecificationsynthesis,
      title={Enchanting Program Specification Synthesis by Large Language Models using Static Analysis and Program Verification}, 
      author={Cheng Wen and Jialun Cao and Jie Su and Zhiwu Xu and Shengchao Qin and Mengda He and Haokun Li and Shing-Chi Cheung and Cong Tian},
      year={2024},
      eprint={2404.00762},
      archivePrefix={arXiv},
      primaryClass={cs.SE},
      url={https://arxiv.org/abs/2404.00762}, 
}

@misc{olausson2024selfrepairsilverbulletcode,
      title={Is Self-Repair a Silver Bullet for Code Generation?}, 
      author={Theo X. Olausson and Jeevana Priya Inala and Chenglong Wang and Jianfeng Gao and Armando Solar-Lezama},
      year={2024},
      eprint={2306.09896},
      archivePrefix={arXiv},
      primaryClass={cs.CL},
      url={https://arxiv.org/abs/2306.09896}, 
}

@misc{tang2026codingagentsfailusers,
      title={How Coding Agents Fail Their Users: A Large-Scale Analysis of Developer-Agent Misalignment in 20,574 Real-World Sessions}, 
      author={Ningzhi Tang and Chaoran Chen and Gelei Xu and Yiyu Shi and Yu Huang and Collin McMillan and Tao Dong and Toby Jia-Jun Li},
      year={2026},
      eprint={2605.29442},
      archivePrefix={arXiv},
      primaryClass={cs.SE},
      url={https://arxiv.org/abs/2605.29442}, 
}

@inproceedings{10.1145/2025113.2025179,
author = {Fraser, Gordon and Arcuri, Andrea},
title = {EvoSuite: automatic test suite generation for object-oriented software},
year = {2011},
isbn = {9781450304436},
publisher = {Association for Computing Machinery},
address = {New York, NY, USA},
url = {https://doi.org/10.1145/2025113.2025179},
doi = {10.1145/2025113.2025179},
booktitle = {Proceedings of the 19th ACM SIGSOFT Symposium and the 13th European Conference on Foundations of Software Engineering},
pages = {416–419},
numpages = {4},
location = {Szeged, Hungary},
series = {ESEC/FSE '11}
}

@ARTICLE{6963470,
  author={Barr, Earl T. and Harman, Mark and McMinn, Phil and Shahbaz, Muzammil and Yoo, Shin},
  journal={IEEE Transactions on Software Engineering}, 
  title={The Oracle Problem in Software Testing: A Survey}, 
  year={2015},
  volume={41},
  number={5},
  pages={507-525},
  doi={10.1109/TSE.2014.2372785}}

@misc{schäfer2023empiricalevaluationusinglarge,
      title={An Empirical Evaluation of Using Large Language Models for Automated Unit Test Generation}, 
      author={Max Schäfer and Sarah Nadi and Aryaz Eghbali and Frank Tip},
      year={2023},
      eprint={2302.06527},
      archivePrefix={arXiv},
      primaryClass={cs.SE},
      url={https://arxiv.org/abs/2302.06527}, 
}

@inproceedings{10.1145/3706598.3713271,
author = {Pickering, Madison and Williams, Helena and Gan, Alison and He, Weijia and Park, Hyojae and Piedrahita Velez, Francisco and Littman, Michael L. and Ur, Blase},
title = {How Humans Communicate Programming Tasks in Natural Language and Implications For End-User Programming with LLMs},
year = {2025},
isbn = {9798400713941},
publisher = {Association for Computing Machinery},
address = {New York, NY, USA},
url = {https://doi.org/10.1145/3706598.3713271},
doi = {10.1145/3706598.3713271},
booktitle = {Proceedings of the 2025 CHI Conference on Human Factors in Computing Systems},
articleno = {875},
numpages = {34},
location = {
},
series = {CHI '25}
}

@inproceedings{10.1145/3613904.3642706,
author = {Nguyen, Sydney and Babe, Hannah McLean and Zi, Yangtian and Guha, Arjun and Anderson, Carolyn Jane and Feldman, Molly Q},
title = {How Beginning Programmers and Code LLMs (Mis)read Each Other},
year = {2024},
isbn = {9798400703300},
publisher = {Association for Computing Machinery},
address = {New York, NY, USA},
url = {https://doi.org/10.1145/3613904.3642706},
doi = {10.1145/3613904.3642706},
booktitle = {Proceedings of the 2024 CHI Conference on Human Factors in Computing Systems},
articleno = {651},
numpages = {26},
location = {Honolulu, HI, USA},
series = {CHI '24}
}

@inproceedings{10.1145/3544548.3580817,
author = {Liu, Michael Xieyang and Sarkar, Advait and Negreanu, Carina and Zorn, Benjamin and Williams, Jack and Toronto, Neil and Gordon, Andrew D.},
title = {“What It Wants Me To Say”: Bridging the Abstraction Gap Between End-User Programmers and Code-Generating Large Language Models},
year = {2023},
isbn = {9781450394215},
publisher = {Association for Computing Machinery},
address = {New York, NY, USA},
url = {https://doi.org/10.1145/3544548.3580817},
doi = {10.1145/3544548.3580817},
booktitle = {Proceedings of the 2023 CHI Conference on Human Factors in Computing Systems},
articleno = {598},
numpages = {31},
location = {Hamburg, Germany},
series = {CHI '23}
}

@inproceedings{10.1145/3519939.3523728,
author = {O'Connor, Liam and Wickstr\"{o}m, Oskar},
title = {Quickstrom: property-based acceptance testing with LTL specifications},
year = {2022},
isbn = {9781450392655},
publisher = {Association for Computing Machinery},
address = {New York, NY, USA},
url = {https://doi.org/10.1145/3519939.3523728},
doi = {10.1145/3519939.3523728},
booktitle = {Proceedings of the 43rd ACM SIGPLAN International Conference on Programming Language Design and Implementation},
pages = {1025–1038},
numpages = {14},
location = {San Diego, CA, USA},
series = {PLDI 2022}
}

@inproceedings{10.1109/ICSE55347.2025.00157,
author = {Bouzenia, Islem and Devanbu, Premkumar and Pradel, Michael},
title = {RepairAgent: An Autonomous, LLM-Based Agent for Program Repair},
year = {2025},
isbn = {9798331505691},
publisher = {IEEE Press},
url = {https://doi.org/10.1109/ICSE55347.2025.00157},
doi = {10.1109/ICSE55347.2025.00157},
booktitle = {Proceedings of the IEEE/ACM 47th International Conference on Software Engineering},
pages = {2188–2200},
numpages = {13},
location = {Ottawa, Ontario, Canada},
series = {ICSE '25}
}

@misc{daplab9patterns,
  title = {9 Critical Failure Patterns of Coding Agents},
  author = {{DAPLab, Columbia University}},
  year = {2026},
  howpublished = {\url{https://daplab.cs.columbia.edu/general/2026/01/08/9-critical-failure-patterns-of-coding-agents.html}},
  note = {Data, Agents, and Processes Lab, Columbia University}
}

@article{10.1007/s10664-025-10614-4,
author = {Tambon, Florian and Moradi-Dakhel, Arghavan and Nikanjam, Amin and Khomh, Foutse and Desmarais, Michel C. and Antoniol, Giuliano},
title = {Bugs in large language models generated code: an empirical study},
year = {2025},
issue_date = {Mar 2025},
publisher = {Kluwer Academic Publishers},
address = {USA},
volume = {30},
number = {3},
issn = {1382-3256},
url = {https://doi.org/10.1007/s10664-025-10614-4},
doi = {10.1007/s10664-025-10614-4},
journal = {Empirical Softw. Engg.},
month = feb,
numpages = {48}
}

@inproceedings{10.1145/2556288.2557420,
author = {Ur, Blase and McManus, Elyse and Pak Yong Ho, Melwyn and Littman, Michael L.},
title = {Practical trigger-action programming in the smart home},
year = {2014},
isbn = {9781450324731},
publisher = {Association for Computing Machinery},
address = {New York, NY, USA},
url = {https://doi.org/10.1145/2556288.2557420},
doi = {10.1145/2556288.2557420},
booktitle = {Proceedings of the SIGCHI Conference on Human Factors in Computing Systems},
pages = {803–812},
numpages = {10},
location = {Toronto, Ontario, Canada},
series = {CHI '14}
}
\begin{acks}
We thank Zhou Yu for her guidance and Haonan Wang for helpful discussions. This research received funding from NSF 2103794, 2312991, 2551201 as well as DAPLab corporate support in the form of funding and/or compute from Amazon, IntellectAI, Infosys, Tidalwave, Veris, shopify, Microsoft, Thinking Machines, Dandy, Perplexity, and Daytona. The views and conclusions presented here are those of the authors and should not be interpreted as representing the official positions of the funding organizations.
\end{acks}
\appendix
\section{Appendix}
\subsection{Constraint.g4 File}
\label{sec:appendix-grammar}
{\small
\begin{verbatim}
grammar Constraint;

// ── entry point ────────────

constraint
    : prob_constraint EOF
    ;

prob_constraint
    : 'P(' logic_expr '|' logic_expr ')' probability_expr
    ;

probability_expr
    : '=' NUMBER
    ;

// ── boolean logic ────────

logic_expr
    : logic_expr OR logic_xor
    | logic_xor
    ;

logic_xor
    : logic_xor XOR logic_term
    | logic_term
    ;

logic_term
    : logic_term AND logic_factor
    | logic_factor
    ;

logic_factor
    : NOT logic_factor
    | '(' logic_expr ')'
    | atom
    ;

// ── atoms ──────────────────

atom
    : write_event
    | user_action
    | system_event
    | persist_event
    | guard
    | literal_bool
    ;

// Three w() forms:
//   w(t)              — existence only
//   w(t, expr)        — value from expr
//   w(t, sources={…}) — value from exact set
write_event
    : 'w(' identifier ')'
    | 'w(' identifier ',' expr ')'
    | 'w(' identifier ',' 
        'sources=' source_set ')'
    ;

source_set
    : '{' source_item (',' source_item)* '}'
    | '{' '}'
    ;

source_item
    : 'r(' identifier ')'
    | 'r(' 'api_result' ')'
    ;

user_action
    : 'A(' identifier ')'
    ;

system_event
    : 'call(' identifier ')'
    | 'call(' identifier ',' expr ')'
    ;

persist_event
    : 'persist(' identifier ')'
    ;

// ── expressions ────────────

guard
    : expr comparator expr
    | expr IN range
    ;

expr
    : expr ('+' | '-') term
    | term
    ;

term
    : term ('*' | '/') factor
    | factor
    ;

factor
    : '(' expr ')'
    | atom_expr
    ;

atom_expr
    : 'r(' identifier ')'
    | 'r(' 'api_result' ')'
    | 'len(' 'r(' identifier ')' ')'
    | 'len(' 'r(' 'api_result' ')' ')'
    | 'status(' identifier ')'
    | 'f(' expr ')'
    | literal
    ;

// ── terminals ──────────────

identifier : IDENTIFIER ;
comparator : '=' | '!=' | '<' | '>' | '<=' | '>=' ;
range      : '[' NUMBER ',' NUMBER ']' | 'D' ;
literal    : NUMBER | STRING | 'null' ;
literal_bool : TRUE | FALSE ;

// ── lexer rules ────────────

NOT   : '¬' | '!' | 'NOT' ;
AND   : '∧' | '&&' | 'AND' ;
OR    : '∨' | '||' | 'OR' ;
XOR   : 'XOR' ;
IN    : 'in' ;
TRUE  : 'true' ;
FALSE : 'false' ;

// IDENTIFIER also accepts a leading '.' so class 
// selectors like `.wishlist-heart` parse as a 
// single token. Downstream layers
// (semantic check, dispatcher) treat the 
// leading-dot form as a CSS class selector to
// bind against dynamic per-instance elements.
IDENTIFIER : '.'? [a-zA-Z][a-zA-Z0-9_-]* ;
NUMBER     : [0-9]+ ('.' [0-9]+)? ;
STRING     : '"' (~["\r\n])* '"' ;

WS : [ \t\r\n]+ -> skip ;
\end{verbatim}
}

\subsection{Parts of CodeQL Primitives}
Full CodeQL queries are on Github at: \url{https://github.com/reyavir/flowcheck}

\textbf{isElementRef}: This predicate is used to identify elements with the specified id, that are initialized in code using document.getElementById(id) or are referenced future uses.
{\scriptsize
\begin{verbatim}
predicate isElementRef(string id, Expr ref) {
  // Direct: `document.getElementById(id)`
  exists(MethodCallExpr mc | mc = ref |
     mc.getMethodName() = "getElementById" and
     mc.getArgument(0).getStringValue() = id
  )
  or
  // Cached: `const x = document.getElementById(id);
  exists(VariableDeclarator decl, MethodCallExpr getEl, Variable v |
     getEl = decl.getInit() and
     getEl.getMethodName() = "getElementById" and
     getEl.getArgument(0).getStringValue() = id and
     v = decl.getBindingPattern().(VarRef).getVariable() and
     ref.(VarRef).getVariable() = v
  )
}
\end{verbatim}
}

\textbf{Writes}
A write to the element is any property assignment on it, or a call to one of a small set of DOM-mutation methods, classList methods, or style methods.

{\scriptsize
\begin{verbatim}
predicate writesElement(string id, AssignExpr write) {
  exists(PropAccess lhs |
    lhs = write.getLhs() and isElementRef(id, lhs.getBase())
  )
}

predicate writesElementVia(string id, MethodCallExpr call) {
  call.getMethodName() = [
    "appendChild", "append", "prepend", "insertBefore",
    "replaceChild", "replaceChildren", "insertAdjacentElement",
    "removeChild", "remove", "setAttribute"
  ] and isElementRef(id, call.getReceiver())
  or
  // element.classList.{add,remove,toggle,replace}(...)
  exists(PropAccess classList |
    classList = call.getReceiver() and
    classList.getPropertyName() = "classList" and
    isElementRef(id, classList.getBase()) and
    call.getMethodName() = ["add", "remove", "toggle", "replace"]
  )
  // (style.setProperty/removeProperty — omitted for space)
}
\end{verbatim}
}

\subsection{LLM Rationalizations of Injected Bugs}
\label{app:llm-rationalizations}
With the most detailed prompt (P3), models often located an injected bug, but then dismissed it by reasoning, rather than flagging it.
Examples from the Amazon app
1. \emph{"qty is clamped to $\geq$ 1, so the qty=0 path is unreachable; not a bug."}
2. \emph{"the qty=99 cap is product-specific per the comment, so missing on other products is likely intentional"}; and 
3. \emph{"USER\_TIER is a const initialized to 'free', so the enterprise branch is dead code, not a defect."}

\end{document}